\documentclass[useAMS,usenatbib]{mn2e}

\usepackage{graphicx}
\usepackage{color}

\def\apj{{ApJ}}

\def\aj{{AJ}}
 
\def\henons{H\'enon's}
\def\henon{H\'enon}
\def\mnras{{MNRAS}}

\def\nbody{$N$-body }
\def\red#1{#1}
\def\gtorder{\mathrel{\raise.3ex\hbox{$>$}\mkern-14mu
             \lower0.6ex\hbox{$\sim$}}}
\def\ltorder{\mathrel{\raise.3ex\hbox{$<$}\mkern-14mu
             \lower0.6ex\hbox{$\sim$}}}

\def\ltorder{\mathrel{\raise.3ex\hbox{$<$}\mkern-14mu
             \lower0.6ex\hbox{$\sim$}}}

\title[Towards an \nbody Model for M4]{Towards an \nbody Model for the
  Globular Cluster M4}
\author[D.C. Heggie]{Douglas C.
Heggie$^{1}$\thanks{E-mail:
d.c.heggie@ed.ac.uk}
\\
$^{1}$University of Edinburgh, School of Mathematics and Maxwell
Institute for Mathematical Sciences, King's
Buildings, Edinburgh EH9 3JZ, UK\\
}
\begin{document}

\date{Accepted \ldots. Received \ldots; in original form \ldots}

\pagerange{\pageref{firstpage}--\pageref{lastpage}} \pubyear{2002}

\maketitle

\label{firstpage}

\begin{abstract}
This paper describes an $N$-body model  for the dynamical evolution of the
nearby globular cluster M4.  The initial conditions, with $N = 484710$
particles, were generated from a
published study of this cluster with a Monte Carlo code.  With the
Monte Carlo code, these initial conditions led, after 12  Gyr of
dynamical and stellar evolution, to a model which resembles M4
in terms of its surface brightness and velocity dispersion profiles,
and its local luminosity function.  Though the \nbody\ model
reported here is marred by 
some errors, its evolution can be compared with that of the published Monte
Carlo model, with a result from the synthetic evolution code EMACSS, and with M4 itself.  
\end{abstract}


\section{Introduction}

Globular star clusters are an obvious target for modelling by \nbody
simulations.  Software techniques have been developed over several decades, and
are well matched to readily available current hardware
\citep{1999PASP..111.1333A, 2012MNRAS.424..545N}.  And yet progress in
applying these methods to the Galactic globular clusters has been very
slow, simply because the computations take such a long time.  Indeed
only two clusters have been adequately modelled, in the sense that
initial conditions have been selected so that the evolved model
matches detailed observational data on the cluster.  These are the
clusters Pal 14 and Pal 4
\citep{2011MNRAS.411.1989Z,2014MNRAS.440.3172Z}.  These are manageable
because the initial particle number ($\ltorder10^5$) is not excessive, and
the initial models are relatively distended (half-mass radius
$\gtorder10$pc), making all time-scales relatively large, and the
computations short.

Without such restrictions (i.e. for the bulk of the Galactic globular
clusters, and certainly the richer and best observed ones) one must
resort to approximate modelling techniques.  One of these is to use
small \nbody models, i.e. models in which the number of stars is much
smaller than that in the actual cluster under study, and to scale the
results appropriately, which means scaling so that the relaxation time
of the model matches that of the real cluster.  An alternative is the
use of a Monte Carlo code
\citep{1998MNRAS.298.1239G,2000ApJ...540..969J,2007ApJ...658.1047F,2013MNRAS.431.2184G}, which yields results
much more quickly than even a scaled \nbody model, but with a
different set of simplifying approximations.  Both of these
approximate methods have been used, for example, to study the
dynamics of stellar-mass black holes in the cluster M22
\citep{2013MNRAS.430L..30S,2014MNRAS.439.2459H}.

Experience shows that finding appropriate initial conditions in such
studies requires the computation of order 50 models or more.
Therefore the speed of the Monte Carlo method, which can compute each
model in a day or two, is a major advantage, except for the sparsest
clusters.  For this reason it has been used in a number of studies of
the richest, closest and best observed clusters
\citep{2008MNRAS.389.1858H,2009MNRAS.395.1173G,2011MNRAS.410.2698G,2014MNRAS.439.2459H},
of which the first was M4.  This is the nearest Galactic globular
cluster, which makes it uniquely favourable for the most delicate
observational programmes \citep[e.g.][]{2013AN....334.1062B}.  

The current mass of M4 has been estimated at about $10^5M_\odot$
\citep{2005ApJS..161..304M}, which seems roughly comparable with Pal 4
and Pal 14.  But two factors make it a harder target for \nbody
modelling:  its smaller half-mass radius (under 1pc initially, according to \citet{2008MNRAS.389.1858H}), which
reduces the time scale of its evolution, and the likelihood that it
has lost a large fraction of its population.  Indeed
the best fitting models which were found by
\citet{2008MNRAS.389.1858H} were those which began with almost
$5\times10^5$ stars.  Nevertheless the continuing importance of this
object, and the availability of appropriate initial conditions from
the Monte Carlo study, motivated the author to attempt a direct \nbody
simulation.  

The description of this model and results, and comparison with
two approximate models (mainly, the
Monte Carlo model) and observational data on M4 are the main purposes of this paper.  The following
section summarises the initial conditions, and software and hardware
issues, including difficulties and flaws in the model.  Section 3 presents the main results on the dynamical
evolution of the model, and a comparison with the Monte Carlo results.
A comparison with observations of the cluster itself is the topic of
Section 4.  The final
section picks out some highlights, and sums up.

\section{Description of the simulation}

\subsection{Initial conditions, software and hardware}

Initial conditions were generated from those of a Monte Carlo model
described by \citet[][Tables 1 and 3]{2008MNRAS.389.1858H}.  Briefly,
the initial model is a Plummer model, with initial tidal and half-mass
radii of 35.0 and 0.58 pc, respectively.  The tidal field was that of
a point-mass galaxy.  The initial number of
objects (single stars plus binaries) was 453000, of which 7\% were
binaries.  Thus the initial number of stars was 484710.  The initial
mass function for single stars was a continuous, two-part power law in
the range $0.1M_\odot < m < 50M_\odot$, with power law indices of 0.9
and 2.3 (Salpeter) below and above a break mass of $0.5M_\odot$.  The
binary mass distribution was drawn from
\citet[][eq.1]{1991MNRAS.251..293K}, with a uniform mass ratio between
components, restricted to  component masses in the
same range as for single stars.  A uniform distribution for the log of
the semi-major axis in the range from 50AU down to twice the sum of
the radii of the components was used.  Eccentricities were thermal,
with ``eigenevolution'' \citep{Kr1995}.  The stellar metallicity was
$Z = 0.002$.  As a result of these choices the  units of mass
and time (\henon\ units\footnote{See \citet{1971Ap&SS..14..151H}.
  Also known as \nbody\ units.}) were, respectively, $350044M_\odot$ (the initial total mass)
and $0.01655$Myr, reflecting the short initial crossing time of the model.

The initial conditions of the Monte Carlo model are incomplete for
\nbody purposes (e.g. the initial position of a star is specified only
as far as the radius), and complete \nbody initial conditions were
generated using the obvious distributions.  \red{For single stars and the
barycentres of binaries, the spherical polar angles of the position are
distributed with probability density function $f(\theta,\phi) =
\sin\theta/(4\pi)$.  Given the magnitude of the radial and transverse components of the
velocity (of a single star or binary barycentre), the sign of the
radial velocity is chosen with probability $0.5$ each, and the
direction of the transverse component is specified by an angle
$\psi$ (in
a plane perpendicular to the radius vector) with probability
density $f(\psi) = 1/(2\pi)$.  For each binary the semi-major axis,
$a$, and eccentricity, $e$, are known from the Monte Carlo data.  The
three standard angles specifying the orientation of the binary orbit
(i.e. the orbit for the relative motion of the two components) is
chosen from $f(\omega,\Omega,i) = \sin i/(8\pi^2)$, and the initial
mean anomaly has probability density $f(M) = 1/(2\pi)$.}

The simulation was run on two virtually identical devices (fermi0 and
fermi1, which were used alternately according to availability)
at Edinburgh Parallel Computing Centre.  Each is equipped with
$4\times6$ Intel Xeon X5650 cores at 2.67GHz, and 4 Nvidia Tesla C2050
GPUs, though, as these are shared machines, only 12 cores and 2 cards
were used for this simulation.

The simulation was carried out with the code NBODY6 in a form which
exploits GPUs \citep{2012MNRAS.424..545N}.  As this code is
continually evolving and improving, however, several different
versions were adopted during the long course of the run.  In some
cases the version was changed in order to progress over a particularly
recalcitrant problem in the simulation.  

Output  was mainly confined to a quite full version of the standard
output of NBODY6, sampled every \henon\ time unit initially, and then
more infrequently as the simulation speeded up.  A complete dump of
the calculation, from which a restart would be possible, was taken
almost daily.  \red{Particle lists extracted from these dumps are available
at http://datashare.is.ed.ac.uk/handle/10283/618\ .}

\subsection{Progress of the simulation}

\begin{figure}
\hspace*{-0.7cm}  \includegraphics[scale=0.95,trim= 40 0 0 0,clip=true]{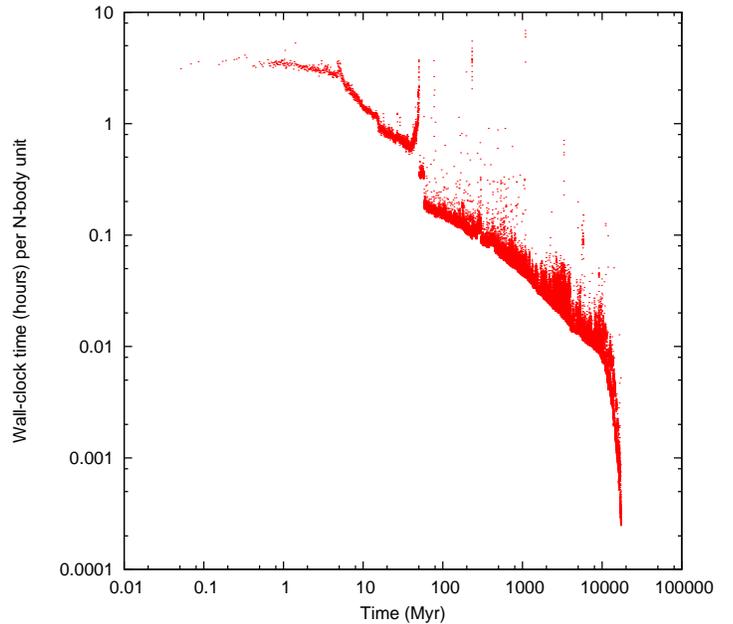}
  \caption{Wall-clock time in hours per \henon\ time unit, as a function
    of simulated evolution time in Myr.}\label{fig:dwt}
\end{figure}

The simulation began in October 2010 and was finally completed after 2
years and 8 months in June 2013.  Only about 5 months were
unproductive, because of downtime and the occasional need for
restarts.  The effort devoted to different parts of the simulation
varied enormously, as can be seen from Fig.\ref{fig:dwt}.  To begin
with each \henon\ time unit  took about 3
hours, at which rate the entire simulation would have taken nearly
400 years.  As shall be seen (Sec.\ref{sec:rh}) an expansion of the
system began around 5 Myr, and then the wall-clock time per \henon\
unit began to diminish.  The decrease accelerated in the latter part
of the simulation because of escape.

Between the early and late phases of the simulation only two features
are noteworthy.  At around 50 Myr a sharp deterioration in performance
was noted (see Fig.\ref{fig:dwt}), accompanied by collapse of a core consisting predominantly
of stellar-mass black holes (Fig.\ref{fig:rc-early},which shows the
early evolution of the core radius).  On inspection a decrease was noticed in the
regularisation parameters $R_{cl}$ and $\Delta t_{cl}$
\citep[][p.145]{2003gnbs.book.....A}, caused by a specific choice of
a certain NBODY6 input parameter\footnote{$KZ(16) = 1$}.  When corrected, the rate of
progress improved by about a factor of 10.  The second feature,
visible above and to the right of the ``main sequence'' of points in
Fig.\ref{fig:dwt}, is a number of phases of poor performance caused by
``difficult''  few-body interactions, when the time taken per \henon\
time unit can increase by as much as a factor of 100.

\subsection{Anomalies}\label{sec:anomalies}

\subsubsection{Roche-lobe treatment}

During the course of the run, two flaws emerged which have a
significant effect on some of the statistical results of interest.
The first concerns binaries, and resulted from a recalcitrant error in
the treatment of one case of  Roche-lobe overflow.  Different
treatments of this process are available in NBODY6, and again a change
of one input parameter\footnote{$KZ(34)$ from 1 to 2} cured the
problem.  However, this also apparently led to a slight decrease in the
binary fraction, from about 4.82\% to about 4.79\%.  This is invisible
on the scale of Fig.\ref{fig:fb}, where the binary fraction is plotted.

\subsubsection{``Anomalous'' black holes}\label{sec:bhanomaly}

Much more serious was a flaw which was noticed when, just before 4 Gyr,
the total bound mass of the model decreased by several hundred solar
masses  
between successive output times.  After some time this was traced to
an unfortunate choice of one of the starting parameters in
NBODY6\footnote{$KZ(28) = 0$, which switches off gravitational
radiation and causes the code to enter routines BRAKE and BRAKE3, where the radii of
black holes and neutron stars are artificially enlarged, leading to enhanced
coalescence.  The NBODY6 community were alerted to the issue on 20
Sep 2011, though the run retained the old value.}.  The result of this flaw was to artificially enhance the
rate of collisions of neutron stars and  black holes with other
objects, and hence (in particular) it led to the build-up of the
mass of individual black holes.  Its effects, which are discussed further below (Sections
\ref{sec:rh},  \ref{sec:core} and \ref{sec:bh}), are particularly
noticeable in the measurements  of  the core radius.  

A list of the detected anomalous black
holes is given in Table \ref{tab:abh}, and for brevity these objects
are referred to henceforth as ABH1, etc,
using the identifier in this table.  
In most
of the early cases each object was involved in only one event, at the
time stated.  All except the last, which was
still present when the simulation ended, presumably escaped.     It
may seem surprising that such a massive object as ABH23 can escape
dynamically, but it was not the only anomalous black hole present at
the time, and interactions between these two objects and normal black
holes may be responsible.  Unfortunately no details of such
interactions were stored.

\begin{table}
  \caption{Detected ``anomalous" black holes}
  
  \begin{tabular}{rrrr}\label{tab:abh}
    Identifier& First detection  & Last detection & Final Mass
   \\
   &  (Myr) &  (Myr)&     ($M_\odot$)\\
1&106.2&106.2&33.4\\
2&106.5&106.5&37.9\\
3&107.1&107.1&38.8\\
4&107.2&107.2&38.6\\
5&206.9&206.9&38.6\\
6&207.2&207.2&38.9\\
7&207.3&207.3&38.9\\
8&207.4&317.4&110.3\\
9&306.4&306.4&38.6\\
10&371.8&371.8&14.3\\
11&405.4&405.4&33.3\\
12&406.2&406.2&34.5\\
13&505.5&505.5&37.8\\
14&506.2&506.2&38.2\\
15&507.0&606.2&58.3\\
16&663.4&663.4&19.8\\
17&704.6&704.6&29.0\\
18&705.1&705.1&38.4\\
19&705.5&705.5&38.8\\
20&706.5&706.5&38.5\\
21&814.1&814.1&19.6\\
22&836.0&1213.7&189.0\\
23&1213.7&3957.0&514.5\\
24&3946.8&5512.8&155.8\\
25&6188.8&9908.7&66.9\\
26&10453.9&10987.9&22.0\\
27&10686.9&11843.3&26.8\\
28&12478.0&12670.0&11.4\\
29&13376.8&19550.6&114.6
  \end{tabular}
\end{table}

\section{Results of the model}\label{sec:results}

In this description of the results a comparison is given, where
possible, with the Monte Carlo simulation of
\citet{2008MNRAS.389.1858H}, though  it is worth mentioning here that the
Monte Carlo
code has been substantially developed since the time of that  
study \citep[see][]{2013MNRAS.431.2184G}.   A comparison is also
presented with the
synthetic cluster evolution code EMACSS
\citep{2014MNRAS.442.1265A}\footnote{Version 3.10 downloaded from \par\noindent https://github.com/emacss/emacss on 12 May
  2014.  The command line used was 
emacss -N 453000 -r 0.58 -s 1  -d 2.25 -v 220  -t 18000 .  This gives
a model which starts with the same total number of objects as the
Monte-Carlo and \nbody models, and the same half-mass and tidal radii.}.

\subsection{Global evolution}\label{sec:global}

\subsubsection{Total mass}\label{sec:mass}

\begin{figure}
\hspace*{-0.7cm}  \includegraphics[scale=0.95,trim= 40 0 0 0,clip=true]{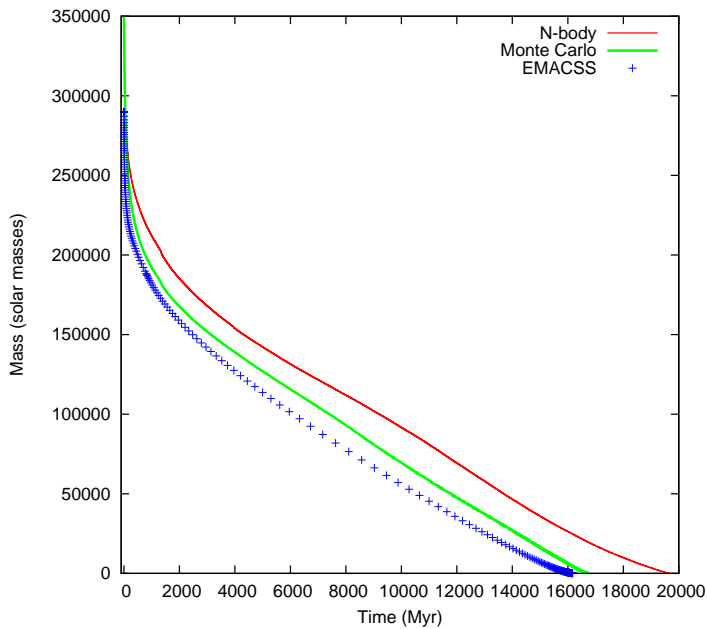}
  \caption{Evolution of the total mass for three dynamical
    evolutionary methods applied to the initial conditions for M4.}
\label{fig:mass}
\end{figure}

The evolution of the mass within the tidal radius is shown in
Fig.\ref{fig:mass}.  All models exhibit the familiar two stages of
mass loss: first, a rapid loss through winds and supernovae (mainly), and,
second, a gentler loss due to escape across the tidal boundary.  

In the first stage, it might be thought surprising that the Monte Carlo model loses mass
more rapidly than the \nbody\ model, as both codes 
 use the
same packages for stellar evolution
\citep{2000MNRAS.315..543H,2002MNRAS.329..897H}.  But the treatment of
escape in the Monte Carlo model is deliberately generous:  the escape
energy is lowered a little in order to match in a crude way the
complicated physics of escape in a tidal field
\citep[][Sec.2.2.2]{2008MNRAS.388..429G}.  While this gives a better
match in general to the overall
lifetime of star clusters it does produce an early burst of escapers,
the effects of which are seen here.  

The reason why EMACSS also lies below the \nbody\ result is
different: it is not possible to adjust the initial mean mass in
EMACSS, which is based on an assumed initial mass function which is
different from that adopted in the other two codes.  In fact if, as is
done here, one arranges for EMACSS to start with the same number of
objects, then the initial total mass is deficient, as one can see in
Fig.\ref{fig:mass}.  (The curve for the \nbody\ model is overprinted
by that for the Monte Carlo model, but the initial total mass in these
two models is in fact the same: $350044M_\odot$.)

In the second stage all three  models lose mass at a very
comparable rate.  Up to a point it is slightly surprising that the
Monte Carlo model agrees so well, as one of the significant
improvements that have been made to the method {\sl since} the time of the
M4 study is exactly in the treatment of escape.   It is gratifying that
the mass loss rate in EMACSS agrees well with the mass loss rate in this
\nbody\ model, as this is not one of the models against which it was calibrated.

\subsubsection{Half-mass radius}\label{sec:rh}

\begin{figure}
\hspace{-0.5cm}  
\includegraphics[scale=0.95,trim= 50 0 0 0,clip=true]{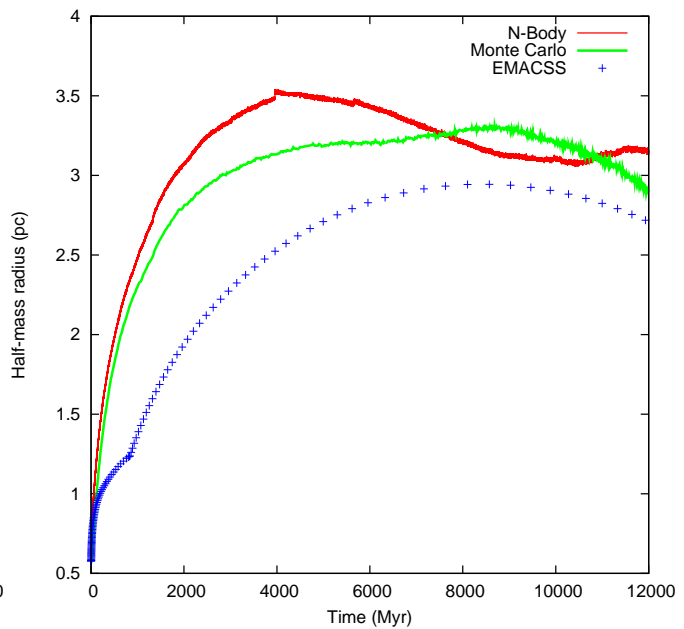}
    \caption{Evolution of the three-dimensional half-mass radius for
    three models of M4.  Incidentally, an effect of the escape of
    ABH23 (Sec \ref{sec:bhanomaly}) is just visible shortly before 4 Gyr in
    the \nbody\ model.}\label{fig:rh}
\end{figure}

The other important global parameter is the half-mass radius
(Fig.\ref{fig:rh}).  This figure may help to explain the shorter lifetime of the
EMACSS model, which expands  too little and too slowly.  This
keeps its dynamical timescales smaller than in the other two models,
which accelerates its dissolution.

In much the same way, the somewhat greater expansion of the
\nbody\ model may help to explain its greater longevity compared with
the Monte Carlo model, though the
initial stages of the rise in the two models are quite comparable.
The subsequent features look dissimilar, but the dissimilarity can be understood in part
as a result of the different evolutionary timescales of the models.
It will be seen later (Sec.\ref{sec:core}) that the slight rise (at about 9  Gyr in the Monte Carlo model and about 12
 Gyr in the \nbody model) is associated with (second) core collapse.
Before that, the  rate of expansion of the half-mass radius
decreases from its initial value to about 0 in the Monte Carlo model, and even reverses in
the \nbody\ model.

There is no evidence that the presence of anomalous black holes
affects the evolution of the half-mass radius, even though they may
lead to more vigorous binary formation, hardening and heating.
According to ``\henons\ Principle'' \citep{1975IAUS...69..133H} the
core adjusts so that the rate of heating becomes equal to that
required by self-similar evolution of the whole system, and that in
turn is determined by global two-body relaxation.  (This is what is
often referred to as ``balanced evolution''.)  Thus the rate of
expansion of the system need not be determined by the efficiency of
the heating mechanism.  

\subsection{Evolution of the core}\label{sec:core}

\begin{figure}
  
\hspace*{-0.5cm}\includegraphics[scale=0.95,trim= 50 0 0 0,clip=true]{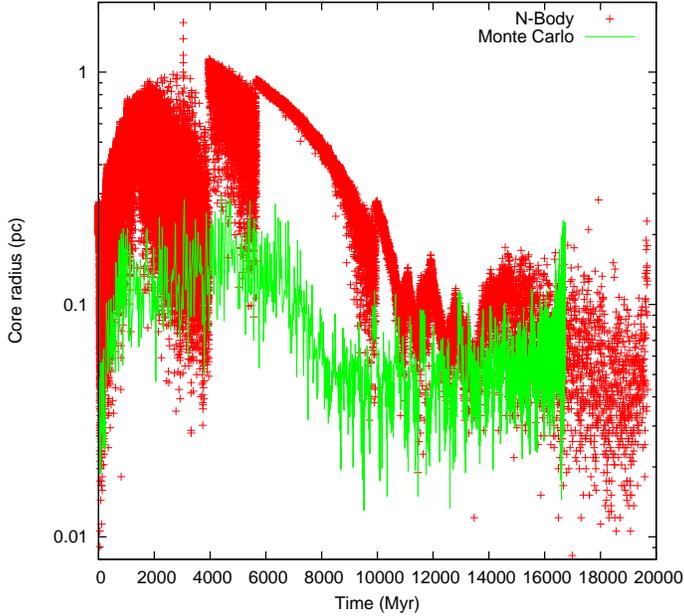}
  \caption{Evolution of core radii for the Monte Carlo and
    \nbody\ models of M4.  Note the effect on the \nbody\ data of the
    escape of ABH23 and ABH24 at about 4 and 6  Gyr, respectively.
    Much of the spread in the \nbody\ values is caused by their
    orbital motion, but other mechanisms contribute noticeably in the
    first  Gyr; see Fig.\ref{fig:rc-early}.}\label{fig:rc}
  \end{figure}

Some complications in the evolution of the core have already been
hinted at in Sections \ref{sec:bhanomaly} and \ref{sec:rh}, and
are visible in the raw results (Fig.\ref{fig:rc}), as will be discussed.  But a
further complication is the definition of ``core radius'', which is
different for the two models.  For the \nbody\ model it is based
entirely on the density distribution, and is estimated along the lines
of \citet{1985ApJ...298...80C} as the rms distance from the density
centre, weighted by a power of the local density.  For the Monte Carlo
model the estimate depends on the velocity dispersion and the central
density, along the lines of \citet{1966AJ.....71...64K}.

The presence
of anomalously massive black holes in the \nbody\ model
(Sec.\ref{sec:bhanomaly}) has a dramatic effect on the data.  Their motion in the
core is responsible for much of the {\sl spread} in values immediately before the loss
of ABH23  just before 4  Gyr.  This spread, however, is nothing more
than a shortcoming of the definition of the core radius, which is
sensitive to the orbital motion of one or more particularly massive bodies.
After about 4  Gyr the absence of ABH23 changes the amplitude of the
spread, but ABH24 causes a continuing but reduced spread in values
until it too escapes at about 6  Gyr.  Despite the spread, the upper envelope of the data shows a clear rise up to about 4  Gyr.  While it is tempting to hold the
anomalous black holes responsible for the rapidity of the rise (as well as for the
spread of values, as mentioned above), 
there is no evidence that the rise would have been any slower with a
normal black hole population.   
Furthermore, despite their continued presence, the rise to 
about 4  Gyr is followed by a steady fall
until about 13  Gyr.  These trends match those in the half-mass radius
quite well chronologically.  Similar remarks (about the trends) may be
made about the Monte Carlo model, which suggests that the role of
anomalous black holes is at most quantitative.

The end of the decreasing phase
in both models is what is referred to in this paper as ``(second) core
collapse''.   
After this point there are no such clear trends, but in
the Monte Carlo data one can (with care) discern brief, quasi-periodic episodes of small core
radius, of which there are about eight between 10 and 14  Gyr.  A
similar phenomenon was noticed in the Monte Carlo model of the globular cluster
NGC 6397 discussed by \citet{2009MNRAS.395.1173G}, and was also visible in an
\nbody\ model of that cluster which covered only a short period around
an age of 12  Gyr and used initial conditions generated from the
evolved Monte Carlo model \citep{2009MNRAS.397L..46H}.

\begin{figure}
  
\hspace*{-0.5cm}\includegraphics[scale=0.95,trim= 50 0 0 0,clip=true]{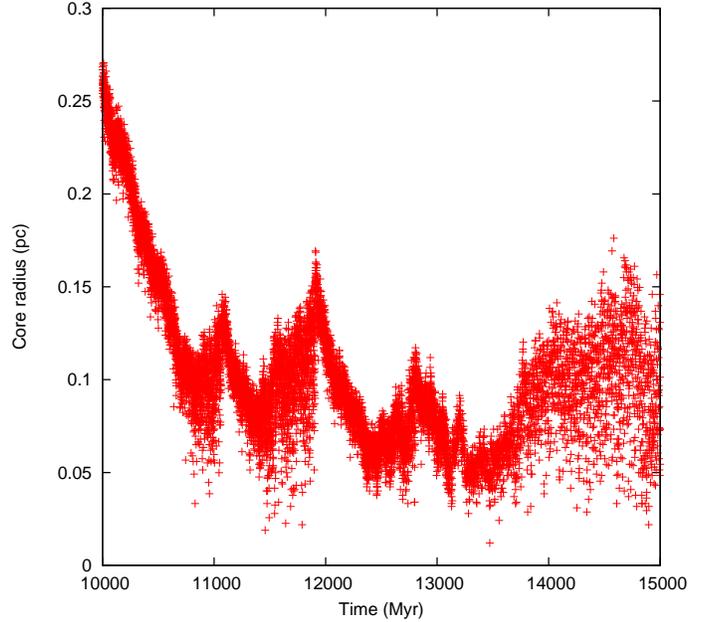}
  \caption{Core radius of the \nbody\ model from 10 to 15  Gyr.}
  \label{fig:nbody-rc}
\end{figure}

Core evolution in the \nbody\ model after second core collapse is
obscured by the Monte Carlo data in Fig.\ref{fig:rc}, and so it is
presented alone in Fig.\ref{fig:nbody-rc}, which begins just after the
escape of ABH25.  Second core collapse is
quite distinct near 11  Gyr.  After about 14  Gyr the spread in values
caused by the growth of ABH29 obscures any detail, but between 11 and
14  Gyr there are clear oscillations on a range of timescales and
amplitudes.  These are not simply caused by the orbital motions of
individual black holes, the crossing time of the whole cluster being
only 0.99 Myr at 12.5  Gyr, and about 0.02 Myr for the core.   Three
anomalous black holes (ABH26--28) escape during the evolution shown in
Fig.\ref{fig:nbody-rc}, but the times of their departure (Table
\ref{tab:abh}) do not seem to be
obviously related to any of the features in the plot, and in any case
their masses would not be untypical of stellar-mass black holes.  The 
oscillations resemble those observed in the aforementioned \nbody\ model of NGC6397
\citep{2009MNRAS.397L..46H}.

\begin{figure}
  
\hspace*{-0.5cm}\includegraphics[scale=0.95,trim= 50 0 0 0,clip=true]{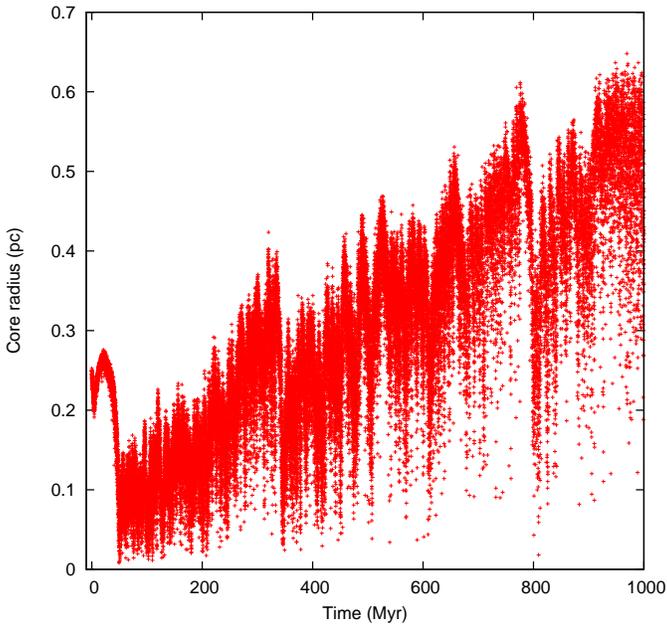}
  \caption{Core radius of the \nbody\ model up to 1  Gyr.}
  \label{fig:rc-early}
\end{figure}

The early evolution of the core radius in the \nbody\ model is even
more curious (Fig.\ref{fig:rc-early}).  There is a brief contraction,
just discernible at the left, until stellar evolution begins at about
5 Myr.  The resulting loss of mass causes a rise, but this weakens, and the
evolution is reversed by continuing mass segregation.  The core,
increasingly dominated by black holes, passes through (``first'') core collapse at
about 50 Myr, and thereafter exhibits post-collapse expansion.  But
overlying the expansionary trend is a long sequence of deep
oscillations.  As with those shown in Fig.\ref{fig:nbody-rc}, these
cannot be caused by orbital motions of individual black holes, the
crossing time at 500 Myr being about 0.04 Myr in the core.  The escape
of anomalous black holes is also unlikely to be responsible, as most
oscillations cannot be associated with these times (Table
\ref{tab:abh}).  A plausible explanation of these oscillations is that
they are gravothermal.

\subsection{Evolution of populations}

\subsubsection{Binaries}

\begin{figure}
  
\hspace*{-0.3cm}\includegraphics[scale=0.95,trim= 50 0 0 0,clip=true]{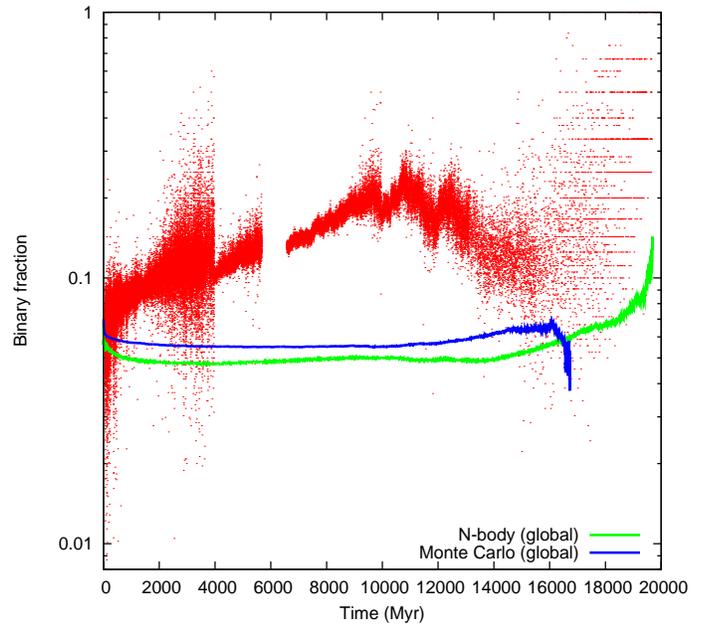}
  \caption{Evolution of the global binary fraction and, for the
    \nbody\ model, the fraction in the core \red{(dots)}.  For this plot, data was
    lost around 6 Gyr.  Core data is not readily available for the Monte Carlo model.
}
\label{fig:fb}  
\end{figure}

\red{After the rapid erosion of soft binaries near the start,} both the \nbody and Monte Carlo models exhibit rather little evolution
of the binary fraction until the end of each model (Fig.\ref{fig:fb}).
For most of the evolution the values in the \nbody\ model are in the range 0.047--0.050.
In the core the \nbody\ data generally exhibit large fluctuations
because of large variations in the estimate of the core radius caused by anomalous
black holes (cf. the discussion in the second paragraph of Sec.\ref{sec:core}).  Still, it is clear that high values, of order 0.2,
prevail at ages in the range 10--12  Gyr.  Note that these are binary
fractions in the three-dimensional core, not the projected core, in
which the binary fraction is 0.046.   Similarly, they include all binaries, including binaries with remnant
components.  If these are excluded, the global binary fraction reduces
to 0.035.  This is much lower than the range of about 0.10--0.15 reported in a recent
study of photometric offset binaries in M4
\citep{2012A&A...540A..16M}.  The primordial value is the same as in
the Monte Carlo model, and at the time when the latter was developed,
the binary fraction was not well constrained \citep[Table 2 in][]{2008MNRAS.389.1858H}.


\subsubsection{Degenerate remnants}\label{sec:bh}

\begin{figure} 
\hspace*{-0.7cm}  \includegraphics[scale=0.95,trim= 40 0 0 0,clip=true]{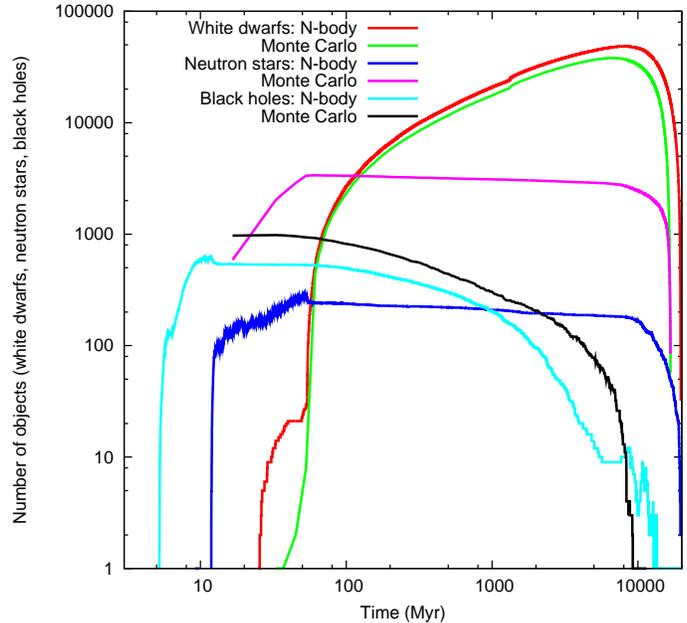}
    \caption{Numbers of white dwarfs, neutron stars and black holes in the \nbody\ and Monte Carlo
    models.}\label{fig:nwd}
\end{figure}

The numbers of white dwarfs in the \nbody\ and Monte Carlo models are
shown in Fig.\ref{fig:nwd}.  After about 50 Myr the differences merely reflect the
different lifetimes and masses of the two models
(Sec.\ref{sec:mass}).  Before that the differences are caused by the
fact that stellar evolution is not updated as frequently in the Monte
Carlo model, because the overall time step is much longer (than in the
\nbody\ model).

The numbers of neutron stars in the two models, also plotted in
Fig.\ref{fig:nwd}, are quite different, but only because no natal
kicks were applied to neutron stars in the Monte Carlo model.  The
numbers in the \nbody\ model, where kicks with a 1-dimensional
dispersion of 190 km/s were applied, are smaller by a factor of approximately
0.07, which may be taken as an evaluation of the retention factor for this model.  

Most problematic is the evolution of the number of black holes
(Fig.\ref{fig:nwd} again).  Initially the numbers in the Monte Carlo
model are higher, because of the absence of natal kicks, though the
cadence of the Monte Carlo time steps 
is hardly adequate at this time.  In the \nbody\ model, black
holes were formed with natal kicks, except in  cases akin to complete
``fallback'' or ``reimplosion'' \citep{1995ApJS..101..181W}, as
determined by the recipes of \citet{2000MNRAS.315..543H}.  The
subsequent evolution of the black hole population is greatly affected
by the ``anomalous'' black holes reported in Sec.\ref{sec:bhanomaly},
but it is possible to argue that the numbers are not greatly affected,
in the following manner.  \citet{2013MNRAS.432.2779B} argued that the
rate at which black holes are lost by escape, in particular the rate
of mass loss, is determined by the overall evolution of the cluster
(mainly, the expansion of the half-mass radius).  In the formation of
an anomalous black hole a certain number of black holes are removed
from the system, but when the ``anomalous'' black hole eventually
escapes, it carries off all the mass of the black holes which
contributed to it.  The effect, on both the mass and number of black
holes, may well be comparable with what would have happened had the
evolution of black holes been calculated correctly.  At any rate it
can be seen that the number of black holes reduces to zero (Monte Carlo)
or one ($N$-body) at about the time of second core collapse, a fact which is
understandable from the theory of \citet{2013MNRAS.432.2779B}; \red{and} this
\red{time} is later in
the \nbody model than in the Monte Carlo model (Sec.\ref{sec:core}).

\subsubsection{Collisions, etc}

\begin{figure}
\hspace*{-0.3cm}  \includegraphics[scale=0.95,trim= 50 0 0 0,clip=true]{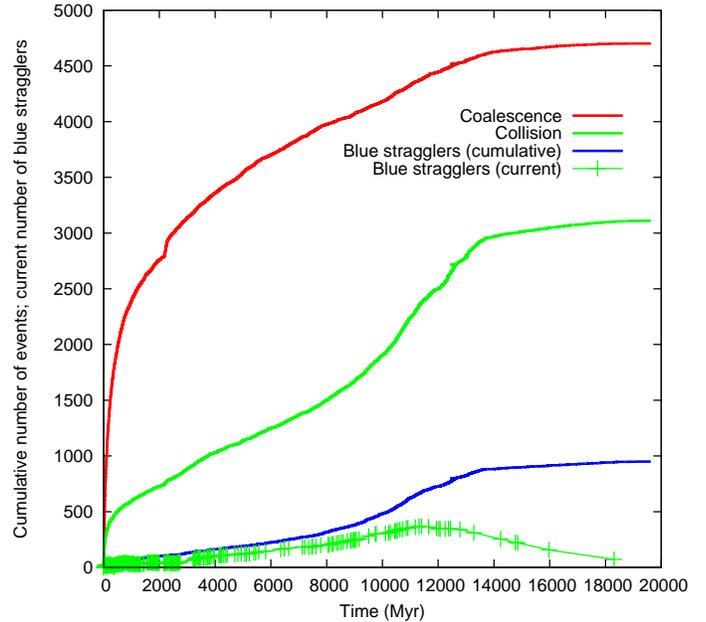}
  \caption{Numbers of collisions, coalescences, and events leading to
    the formation of blue stragglers; and the current number of blue
    stragglers.  The last of these is based on much more sparsely
    sampled data.}\label{fig:ncoll}
  
\end{figure}

The cumulative number of  collisions is plotted in
Fig.\ref{fig:ncoll}, which shows substantial increases around the time of both
first and second core collapse.  Collisions arise in two-body and
higher multiple encounters, and are distinguished from {\sl
  coalescence} following Roche-lobe mass transfer.  The cumulative numbers of
these are also plotted, along with one of the possible outcomes of
these events, i.e. the formation of a blue straggler.  All three curves flatten out as the model loses almost all
its stars towards the end of its life.  Note, however,
that these are the cumulative numbers formed, and not the current
number, which is diminished by escape.  As an example, the figure also
shows the current number of blue stragglers (defined as main sequence
stars with a mass at least 1.02 times the turnoff mass).

\section{Comparison with observations}

\subsection{Surface brightness profile}

\begin{figure}
  \includegraphics[scale=0.95,trim= 50 0 0 0,clip=true]{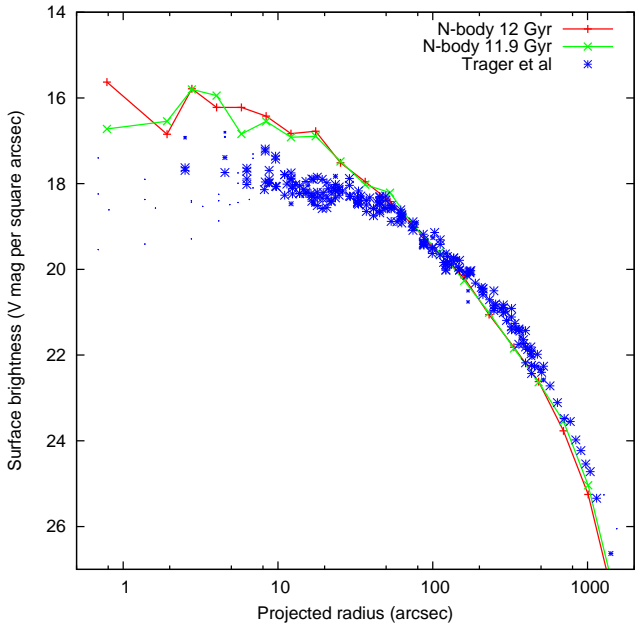}
  \caption{Surface brightness of the model at 12  Gyr, compared with
    the observational profile of \citet{Tr1995}.  The size of points
    in the latter is determined by the weight which those authors
    assigned to different data; note in particular the scatter of
    small points at small radii.  The model profile at 11.9 Gyr is
    also shown.}\label{fig:sb}
  \end{figure}

For the purposes of comparison with observations, the  state
of the model was extracted from the complete dump file at a simulated
time of 12  Gyr, which is the age for M4 that has been  assumed
in this study,  as in \citet{2008MNRAS.389.1858H}.  To compute
the surface brightness, the positions were first projected along the
three coordinate axes\footnote{While more than three projections could be used,
  there would then be an
increased risk of bias.  If, for example, most bright stars happen to
be close to apocentre, the core would tend to be too dim.}, and the location of the centre of the cluster
was taken to lie at the median values of the coordinates.  The surface
brightness was computed in a series of annuli, uniform in the
logarithm of the projected radius, and averaged over the
three projections.    Since the dump file includes only
the luminosity and radius of each star, a simple bolometric correction
was applied \citep{1998JRASC..92...36R}, but it was also verified that
$V$-magnitudes, calculated according to recipes kindly provided by
J. Hurley, produced an almost indistinguishable surface brightness profile.  Extinction was applied to
the resulting surface brightness, and the result is compared with the
observational profile from \citet{Tr1995} in Fig.\ref{fig:sb}.

While the overall size of the model is acceptable, in detail the
profile is too concentrated.  The central surface brightness is too
bright by about two magnitudes, and the profile is too dim beyond
about 100 arcsec, i.e. not far outside the observational {\sl core}
radius of about 70 arcsec \citep[][2010 revision]{Ha1996}.  In almost
every respect the Monte Carlo model did better \citep[Fig.1 in][]{2008MNRAS.389.1858H}.  The Monte Carlo code (now called MOCCA)
has been improved considerably since the time of that study
\citep{2013MNRAS.431.2184G}, and it is possible that the newer version
would share the problems of the \nbody\ model, indicating merely that
the initial conditions were not optimal for either the \nbody\ code or
MOCCA.  But the problem does not appear to lie with the model at 12
 Gyr.  Though  the dynamical core radius appeared to be
somewhat larger at 11.9  Gyr 
(Fig.\ref{fig:rc}), suggesting that 
the \nbody\ model would appear less concentrated, the 
surface brightness profile was little better (Fig.\ref{fig:sb}).

A possible reason for the  brighter, smaller core in the
\nbody\ model is the relative paucity of black holes and neutron
stars.  It is known \citep{2004ApJ...608L..25M,2008MNRAS.386...65M}
that a large retained population of black holes drives an expansion of
the core radius, and it may be that a large retained population of
neutron stars could enhance this effect slightly.  Incidentally,  the central mass to light ratio
(in projection) is $M/L_V \simeq 1.4$ in solar units, 
while the global value is 2.16.

\subsection{Velocity dispersion profile}

\begin{figure}
  \includegraphics[scale=.95,trim= 50 0 0 0,clip=true]{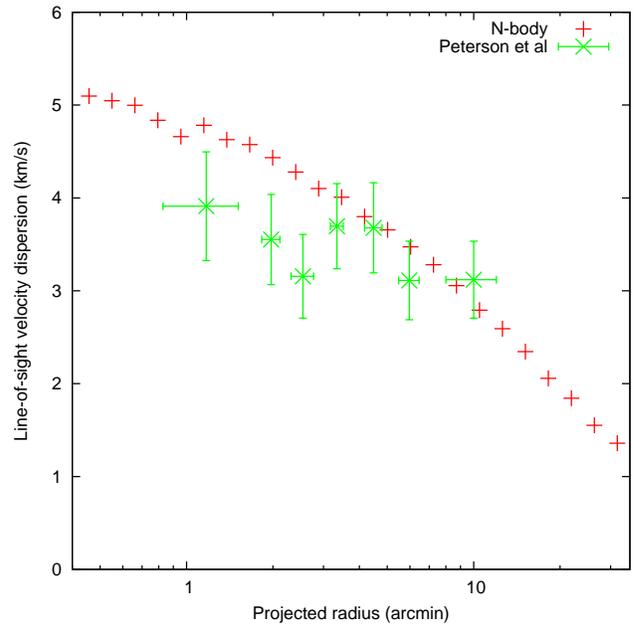}
  \caption{Line-of-site velocity dispersion in the \nbody\ model,
    compared with the observational data of \citet{Pe1995}.}\label{fig:vdp}
  
\end{figure}

To construct the velocity dispersion profile (Fig.\ref{fig:vdp}), an average over three lines of sight (i.e. the three coordinate
axes) is again taken, and  only stars brighter than a limit about 2
magnitudes fainter than turnoff have been used.  For each projection the stars were
binned in projected radius centred on the density centre,  the
mean square line-of-sight velocity was corrected for the mean
l.o.s. velocity of the cluster, and the results were averaged over the
three projections. 

The agreement with observational data \citep{Pe1995} is certainly no
worse than for the surface brightness profile.  The observational data
is confined to the region from about the core radius outward, and
again the problems of the \nbody\ model seem mainly confined to the
region of the core, much as in Fig.\ref{fig:sb}.  At $69300M_\odot$,
the \nbody\ model is somewhat more massive than the Monte Carlo model
($46100M_\odot$) at age 12  Gyr (Fig.\ref{fig:mass}), which itself
provided a satisfactory fit to the observational data, and this is
sufficient quantitatively to account for the poorer results in
Fig.\ref{fig:vdp}.  The fact that the \nbody\ model, with a half-mass
radius of 3.13 pc, is slightly larger than the Monte Carlo model (2.90
pc) makes little difference.

\subsection{Luminosity functions}

\begin{figure}
\includegraphics[scale=.95,trim= 50 0 0 0,clip=true]{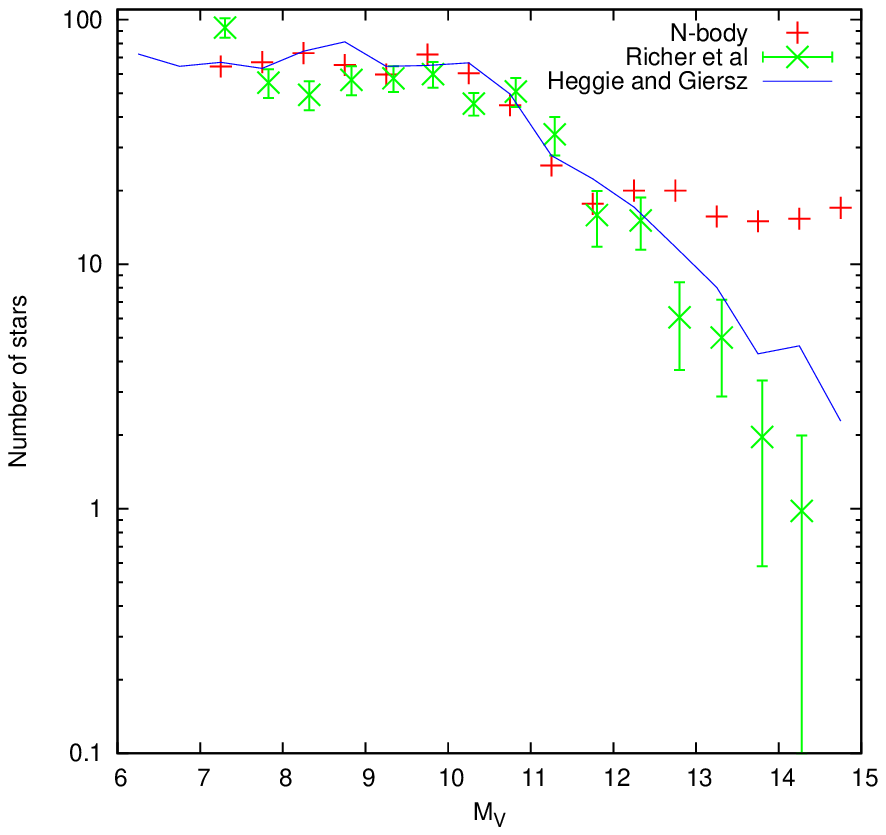}
\includegraphics[scale=.95,trim= 50 0 0 0,clip=true]{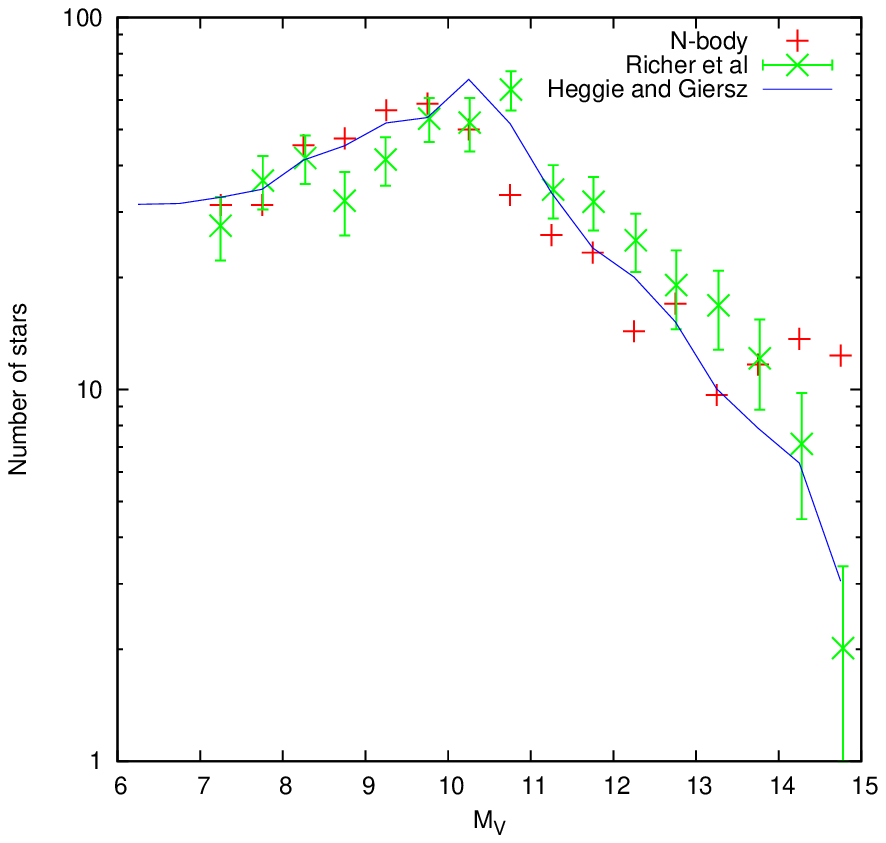}
  \caption{V luminosity functions in annuli with median radii of 0.938
    (top) and 5.028 arcmin (bottom).  The \nbody\ model is compared
    with the observational data of \citet{Ri2004} \red{and the Monte Carlo
    results of \citet{2008MNRAS.389.1858H}}.}\label{fig:lfs}

\end{figure}

Here the \nbody\ model will be compared with the innermost and outermost of the four local
luminosity functions observed by \citet{Ri2004}.  Again the model
results have been obtained by projecting the model along the three
coordinate axes.  In the model, those stars  were counted which lay in two annuli
which have the same area and median radius (after dimensional scaling)
as in the two observational fields.  The results are compared in
Fig.\ref{fig:lfs}.

Since both the observational and \nbody\ data are numbers of stars in a
given area, the Poisson errors will be comparable, even though error
bars are only given for the observational data.  The most noticeable
feature of the inner profile (top) is the great excess of stars at the
faintest magnitudes.  Part of this may perhaps be attributed to
completeness, for which the observational data were not corrected.
According to \citet{richeretal2002}, incompleteness in the {\sl outer}
field is negligible down to $M_V = 15$, though it could presumably be
significant in the inner field.  Another possible explanation is the
effect of anomalous black holes, as \citet{2008ApJ...686..303G} found
that mass segregation was diminished in \nbody models by the presence
of an intermediate-mass black hole.  But in their models the smallest
black hole mass was 0.9\% of the cluster mass, while for the most massive
anomalous black hole in the \nbody\ model of M4 (at least, at any time
since the escape of ABH24 almost 7 Gyr earlier) the fraction is about 0.1\%, which diminishes the
strength of this mechanism.  Whatever the explanation,  the \nbody\ data greatly
exceeds the corresponding result for the Monte Carlo model at faint
magnitudes in the inner field \red{(Fig.\ref{fig:lfs}, top)},
indicating once again considerable
differences between the two models.  Nevertheless in the outer field
(Fig.\ref{fig:lfs}, bottom) the comparison with the observational data
is not very discrepant, except at the faintest magnitudes, and
near the peak of the observational luminosity function: \red{the
  reduced chi-square is 2.2, of which about half is contributed by the
  points at magnitudes 10.75 and 14.75.}

\section{Conclusions and discussion}

\subsection{Conclusions}

In this paper  results are reported of the largest sustained
\nbody\ model of a globular star cluster to date.  It began with $N =
484710$ stars, and took about two years and eight months, using the
code NBODY6 with extensions for GPU support
\citep{2012MNRAS.424..545N}.  The initial conditions were generated
from a Monte Carlo model \citep{2008MNRAS.389.1858H} which, after 12
 Gyr of simulated evolution, resembled the globular star cluster M4 in
respect of its surface brightness and radial (line-of-sight) velocity
dispersion profiles, and its observed luminosity function at two radii.

The evolution of the total mass and half-mass radius in the
\nbody\ and Monte Carlo models are fairly comparable
(Figs.\ref{fig:mass}, \ref{fig:rh}), though the mass of the Monte
Carlo model is systematically smaller, leading to a shorter-lived model.  This may be caused by the
treatment of escape in the Monte Carlo model.  The larger mass of the
\nbody\ model is one factor which may contribute to the moderate
disagreement, at small projected radii,
between the line-of-sight velocity dispersion of the model and the
observational data (Fig.\ref{fig:vdp}).   The evidence of the
luminosity function, when compared with observational results
(Fig.\ref{fig:lfs}), is that some excess mass may be found in the
lower main sequence at small projected radii.   The
most serious disagreement, however,  concerns  the  surface
brightness profile of the core, where the \nbody\ model is too bright
to be consistent with observations (Fig.\ref{fig:sb}).  The most
plausible explanation for this is the relatively small retention
factor of black holes and neutron stars, at least by comparison with
the Monte Carlo model (Fig.\ref{fig:nwd}), where no natal kicks were
applied.

Discussion of the core
radius is complicated by a recurrent error in the \nbody\ model, which
led to spurious coalescence of degenerate remnants, especially black
holes (Table \ref{tab:abh}).  Though it can be argued that this does
not affect the bulk evolution, its effect on the core cannot be so readily
dismissed.  One of these effects is the resulting size of fluctuations in the
estimated core radius when these ``anomalous'' massive black holes were present
(Fig.\ref{fig:rc}).  Nevertheless these anomalous black holes do not
seem to be responsible for some very interesting aspects of core
evolution, especially the oscillations observed in the core radius in
about the first  Gyr of the simulation, and even at times around the
present age of M4 (Figs \ref{fig:nbody-rc},\ref{fig:rc-early}). 

There is rather little evolution of the binary fraction, except in the
core (Fig.\ref{fig:fb}), though the value is too low for consistency
with observational data, because of a poor choice of the initial value.
Though this initial binary fraction was 7\%, \red{almost 5000} of the
binaries led to coalescence as a result of Roche-lobe overflow
(Fig.\ref{fig:ncoll}), which is considerably more than 
the number of
stellar collisions in encounters.   About a thousand of \red{all} these events
were identified by the code as leading to the formation of a blue
straggler. 

\subsection{Discussion}

Evidently the \nbody\ model presented in this paper fails to serve as
a completely satisfactory model of M4.  Faults could lie within
aspects of the dynamical evolution of M4 which are excluded in the
\nbody\ and Monte Carlo models, such as the formation of a second
generation, or conceivably the rotation of the cluster. \red{ The
  choice of a circular Galactic orbit in a point-mass Galactic
  potential was made to facilitate
  comparison with the Monte Carlo results, and implicitly ignores the
  effects of disk and bulge shocking.  Actually, no greater effort would
  have been required to use a more realistic orbit (see, for example,
  \citealt{1999AJ....117.1792D}) and potential.}  But within the \red{self-imposed}
constraints of the modelling underlying this paper, it is worth
considering how a better model might be found, besides some obvious
improvements in the initial conditions, such as the primordial binary
fraction.

Some technical problems of the \nbody\ model are easily avoided,
such as occurrence of the spurious collisions between black
holes which marred the model described in this paper.  But it is clear
that there is one important respect in which the \nbody\ model was
physically more realistic than the Monte Carlo model which generated
the initial conditions, and that is the treatment of natal kicks given
to black holes.  The first step to be taken, then, could be to repeat
the Monte Carlo calculation, but using a similar prescription for
natal kicks as in the \nbody\ model.  If it  turned out that such a
Monte Carlo model  exhibited a similar bright core to that of the
\nbody\ model, it would add significant confirmation of the
reliability of the Monte Carlo model for this kind of work. 
Then it would be profitable to repeat the
exercise of using the Monte Carlo model to find appropriate initial
conditions for M4, especially as the Monte Carlo code (now named
MOCCA) has been improved substantially in the meantime.  Likewise,
ongoing improvements in \nbody\ modelling might then make
it possible to confirm the new Monte Carlo results with
\nbody\ techniques in less time than was taken for the \nbody\ model
described in this paper.

An important point exposed by this discussion is the role played by
the fit to the surface brightness profile.
If the comparison (between the models and with the cluster) had been
restricted to a discussion of the half-mass radius and the mass then
it would have been much less clear how unsatisfactory the
\nbody\ model was.   A second important point is the continuing
central role played by the Monte Carlo method in this field.  Finding
initial conditions requires the computation of many models -- at least
a few dozen.  Even if the run-time of an \nbody\ model could be reduced
from two or three years to two or three months, the time required for
finding appropriate initial conditions for a single cluster would
still be many years.

\section*{Acknowledgements}

Much of the simulation was run at Edinburgh University on the host fermi0, which is supported
by the Centre for Numerical Algorithms and Intelligent Software
(funded by EPSRC grant EP/G036136/1 and the Scottish Funding Council).
I am very grateful to NAIS for the provision and availability of
fermi0.  I thank Sean McGeever, George Beckett and Adrian Jackson, all
of Edinburgh Parallel Computing Centre, for much assistance with the
technical aspects of running on fermi0.  I am indebted to Sverre
Aarseth and Jarrod Hurley for advice on several technical matters,
including the treatment in NBODY6 of natal kicks of black holes.  Mark
Gieles kindly helped with getting the parameters  of the  EMACSS run correct.
I am grateful also to Mirek Giersz and Anna Lisa Varri who generously
gave their time to read and comment on an earlier version of the
paper, and to Hagai Perets for comments about blue stragglers.  \red{I thank
the anonymous referee for a number of useful suggestions, especially
the gentle push to put the data online.}

\bsp

\label{lastpage}

\end{document}